\documentstyle[12pt]{article}

\newcommand{\sect}[1]{\setcounter{equation}{0}\section{#1}}

\newcommand{\EQ}{\begin{equation}}
\newcommand{\EN}{\end{equation}}
\newcommand{\bea}{\begin{eqnarray}}
\newcommand{\ena}{\end{eqnarray}}

\renewcommand{\a}{\alpha}
\renewcommand{\b}{\beta}

\renewcommand{\d}{\delta}

\newcommand{\pa}{\partial}

\newcommand{\G}{\Gamma}

\newcommand{\e}{\epsilon}

\newcommand{\k}{\kappa}

\renewcommand{\l}{\lambda}

\newcommand{\m}{\mu}

\newcommand{\n}{\nu}

\newcommand{\x}{\chi}

\newcommand{\p}{\pi}

\newcommand{\s}{\sigma}

\renewcommand{\t}{\tau}

\begin{document}
 
\topmargin 0pt
\oddsidemargin 5mm
 
\renewcommand{\Im}{{\rm Im}\,}
\newcommand{\NP}[1]{Nucl.\ Phys.\ {\bf #1}}
\newcommand{\PL}[1]{Phys.\ Lett.\ {\bf #1}}
\newcommand{\NC}[1]{Nuovo Cimento {\bf #1}}
\newcommand{\CMP}[1]{Comm.\ Math.\ Phys.\ {\bf #1}}
\newcommand{\PR}[1]{Phys.\ Rev.\ {\bf #1}}
\newcommand{\PRL}[1]{Phys.\ Rev.\ Lett.\ {\bf #1}}
\newcommand{\MPL}[1]{Mod.\ Phys.\ Lett.\ {\bf #1}}
\renewcommand{\thefootnote}{\fnsymbol{footnote}}
\newpage
\begin{titlepage}
\vspace{2cm}
\begin{center}
{\bf{{\large GRAVITATIONAL DRESSING OF $N=2$ SIGMA--MODELS  }}} \\
{\bf{{\large BEYOND LEADING ORDER }}} \\
\vspace{2cm}
{\large S. Penati, A. Santambrogio and D. Zanon} \\
\vspace{.2cm}
{\em Dipartimento di Fisica dell' Universit\`{a} di Milano and} \\
{\em INFN, Sezione di Milano, Via Celoria 16, I-20133 Milano, Italy}\\
\end{center}
\vspace{2cm}
\centerline{{\bf{Abstract}}}
\vspace{.5cm}
We study the $\b$--function of the $N=2$ $\s$--model coupled to
$N=2$ induced supergravity. We compute corrections to first order
in the semiclassical limit, $c \rightarrow -\infty$, beyond 
one--loop in the matter fields. As compared to the corresponding bosonic,
metric $\s$--model calculation, we find new types of contributions
arising from the dilaton coupling automatically
accounted for, once the K\"ahler potential is coupled to $N=2$ supergravity.

\vfill
\noindent
IFUM--550--FT \hfill {February 1997}

\end{titlepage}
\renewcommand{\thefootnote}{\arabic{footnote}}
\setcounter{footnote}{0}
\newpage

\sect{Introduction}
Matter systems which are not conformally invariant, when coupled to
two--dimensional gravity, induce propagating modes of the latter
due to the appearance of the one--loop trace anomaly \cite{b1}. Thus it is of 
interest to study the gravitational back reaction and analyze
how physical quantities are affected by the presence of this
dynamically generated gravitational field. At one--loop order
in the matter fields exact results (i.e. to all orders in the gravitational
coupling) have been obtained for the $\b$--functions of bosonic
\cite{b2}
as well as supersymmetric $N=1,2$ $\s$--models \cite{b3}. More precisely
it has been shown that the effect of induced gravity for the
$N=0,1$ theories is simply to
multiplicatively renormalize the $\b$--functions as follows:
\bea
N=0\qquad &&\b^{(1)}_G =\frac{\k+2}{\k+1} \b^{(1)}_0
\nonumber\\
&&\k+2=\frac{1}{12}[c-13-\sqrt{(1-c)(25-c)}]\nonumber\\
&&~~~~~~~~~\nonumber\\
&&~~~~~~~~~\nonumber\\
N=1\qquad &&\b^{(1)}_G =\frac{\k+\frac{3}{2}}{\k+1} \b^{(1)}_0
\nonumber\\
&&\k+\frac{3}{2}=\frac{1}{8}[c-5-\sqrt{(1-c)(9-c)}]
\ena
where $\k$ is the central charge of the gravitational $SL(2R)$
Kac--Moody algebra.
For the $N=2$ $\s$--model the coupling to supergravity does not 
produce any new divergent contribution so that the one--loop
$\b$--function does not receive a gravitational dressing.
These results indicate that to first order in perturbation theory
for the matter fields, even if the renormalization group trajectories
might be affected (e.g. for the $N=0,1$ theories), the critical points
of the various theories are the same as in the absence of gravity,
since the $\b$--function
is at most rescaled by a multiplicative factor.

The expectation that this result may be universal, valid to all orders
in the matter loops, fails however to be fulfilled at least in the case of the
bosonic $\s$--model. Explicit calculations \cite{b4,b5} 
at two--loop order in the matter
and to leading order in the semiclassical limit, $c\rightarrow -\infty$,
have shown that new structures are produced so that the multiplicative 
renormalization of the $\b$--function is not maintained. Even though 
one would always prefer a simple answer, this result does not come
as a surprise and it indicates that new physics is produced by the 
coupling to a curved quantum spacetime.

It becomes of interest to address the corresponding question for the
supersymmetric theories. The model with $N=1$ supersymmetry is for several
aspects very similar to the bosonic one. On a flat two-dimensional
worldsheet it has been studied at the 
perturbative level using the quantum--background field expansion in
normal coordinates \cite{MF} and the $\b$--function has been computed up to high
orders in the loop expansion. The leading contribution, proportional to the
Ricci tensor, is at one loop, while the next to the leading nonvanishing 
correction is at four loops \cite{b6}. As mentioned above
the coupling to $N=1$ induced supergravity affects multiplicatively the
one--loop $\b$--function. 
We  have not performed the explicit calculation at higher--loop level
but, in a way completely parallel to the bosonic example, 
we expect that due to the interaction with the gravitational field  new 
structures will arise  modifying the flat fixed points of the theory.
Although it might be of interest to know the exact expression of
the gravitational modifications, in a sense no novelties are expected.

The $\s$--model with $N=2$ supersymmetry is singled out already at the one--loop
level being its $\b$--function completely unaffected by the coupling to
supergravity. In this case the interest in studying the situation at
higher perturbative orders in the matter fields is at least two folded: first
there is the question whether this absence of gravitational renormalization
will persist beyond the leading correction.
The other issue that can be studied in the $N=2$ context is the 
influence on the metric $\b$--function of the dilaton coupling.
Indeed in the $N=2$
model the dilaton term is automatically accounted for through the
coupling of the K\"ahler potential to the supergravity fields \cite{b7}: 
although
the classical dilaton coupling vanishes, it reappears and becomes relevant
at the
quantum level. We have found that its presence introduces
significant modifications to the renormalization group trajectories
of the $N=2$ $\s$--model coupled to quantum dynamical $N=2$ 
supergravity. 

Our paper is organized as follows: in the next section we define
the model in $N=2$ superspace and we
briefly summarize the relevant steps which lead to the identification
of the dilaton coupling. In section 3 we present the
quantization of the matter--supergravity coupled system.
We use a formulation of the theory in $d=2-2\e$ dimensions
and follow closely the perturbative approach in Refs. \cite{b10,b5,b8}
to which the reader should refer to for details on the quantization
of the gravitational fields and of the $N=2$ $\s$--model respectively.
Our results are contained in
section 4. We conclude with some final comments.
Notations and useful formulae for K\"ahler manifolds are in Appendix A, 
details of an explicit calculation in Appendix B.

\sect{N=2 matter--supergravity model}

We study the $N=2$ supersymmetric $\s$--model coupled to
$N=2$ supergravity using a superfield formulation.
The complete action can be written in $d=2-2\e$ dimensions
by dimensional reduction from the $N=1$ theory in four
dimensions \cite{b9}
\bea
S&=&S_{SG}+S_{M} \nonumber\\
&=&-2\frac{d-1}{d-2}\k_0^{-2} \int d^d x d^2 \theta d^2 
\bar{\theta}~ E^{-1} 
+\int d^d x d^2 \theta d^2 \bar{\theta}~ 
E^{-1} K(\Phi^{\m}, \bar{\Phi}^{\bar{\m}}) 
\label{action}
\ena
where $E$ is the $N=2$ vielbein superdeterminant,
and $K$ is the K\"ahler potential of the $n$--dimensional complex 
manifold described by $n$ covariantly chiral superfields $\Phi^{\m}$, 
$\bar{\nabla}_{\a} \Phi^{\m}=0$, $\m=1, \cdots ,n$. 
(One could add to this action
a superpotential, a chiral integral term, but due to the 
$N=2$ nonrenormalization theorem
it would play no role in the analysis of the renormalization properties 
of the theory).
 
In two dimensions and in its minimal formulation, the constraints of
$N=2$ supergravity are solved in terms of two prepotentials, 
a real vector superfield $H_{a}$ and a chiral scalar compensator 
$\s$ \cite{GW}. In conformal gauge only the compensator is relevant since $H_a$
contains pure gauge modes. Away from two dimensions $H_a$ cannot be
gauged away completely. However we expect the perturbative 
calculation of the $\b$--function, which is evaluated
in the limit $\e \rightarrow 0$, to be independent of $H_a$.
Indeed the validity of this result  
has been checked explicitly in the bosonic case  up to two 
loops in the matter fields \cite{b5}. 
Dropping the dependence on
 $H_{a}$, with a natural definition of conformal gauge in $d$ 
dimensions \cite{b7}, we write
\EQ
E = e^{\frac{d-2}{2}(\s +\bar{\s})}
\label{1a}
\EN
so that the action for the matter system becomes
\EQ
S_{M} = \int d^dx d^2 \theta d^2 \bar{\theta}~
e^{\e(\s +\bar{\s})} K(\Phi^{\m},\bar{\Phi}^{\bar{\m}})
\label{3}
\EN
Note that defining 
\EQ
\tilde{K}(\Phi^{\m},\bar{\Phi}^{\bar{\m}},\s,\bar{\s}) \equiv
e^{\e(\s+\bar{\s})} K(\Phi^{\m},\bar{\Phi}^{\bar{\m}})
\EN
the action in eq. (\ref{3}) describes a $\s$--model
on a $(n+1)$--dimensional complex manifold with K\"ahler potential 
$\tilde{K}$. 

In order to clarify the role played by the $N=2$ supergravity $\s$
superfield, it is convenient to reexpress the theory in $N=1$ 
superspace.  
To this end we introduce coordinates $\theta_1^{\a} = 
\theta^{\a} + \bar{\theta}^{\a}$,
$\theta_2^{\a} = \theta^{\a} - \bar{\theta}^{\a}$
and corresponding derivatives $D_{1\a}=D_{\a}+\bar{D}_{\a}$, 
$D_{2\a}=D_{\a}-\bar{D}_{\a}$,
and define the $N=1$ projections of the superfields as
\EQ
\Psi^{\m}(\theta_1) \equiv 
\left. \Phi^{\m}(\theta,\bar{\theta}) \right|_{\theta_{2}=0} 
\qquad \qquad~~
\s_1(\theta_1)  \equiv \left. \s(\theta,\bar{\theta}) 
\right|_{\theta_2 =0} 
\EN 
In eq. (\ref{3}) now we integrate on the $\theta_2$ variables 
following Ref. \cite{b7} 
and obtain
\bea
S_{M} &=& \frac14 \int d^dx d^2 \theta_1 (D_2)^2 \left. 
\tilde{K} \right|_{\theta_2 =0} \nonumber \\ 
&~&~~~~~~~~~~~~~\nonumber \\
&=& \frac12 \int d^dx d^2\theta_1 e^{\e(\s_1 + \bar{\s}_1)}
\left[ \frac{\pa^2 K}{\pa \Psi^{\m} \pa \bar{\Psi}^{\bar{\n}}} D_{1\a}
\Psi^{\m} D_1^{\a} \bar{\Psi}^{\bar{\n}} + \right. \nonumber \\
&~~& \left. +\e \left( \frac{\pa K}{\pa \Psi^{\m}} D_{1\a} \Psi^{\m}
D_1^{\a} \bar{\s}_1 + \frac{\pa K}{\pa \bar{\Psi}^{\bar{\m}}} D_{1\a} 
\bar{\Psi}^{\bar{\m}} D_1^{\a} \s_1 \right) + \right.  \nonumber  \\
&~~& \left. + \e^2 K D_{1\a} \s_1 D_1^{\a} \bar{\s}_1 
\right] 
\label{4}
\ena
At this stage,
in terms of the complex structure $J^i_{~j}$ of the K\"ahler manifold, 
we can rewrite the action (\ref{4}) using real coordinates 
$\Psi \pm \bar{\Psi}$ and $\s_1 \pm \bar{\s}_1$ as
\bea
S_{M} &=& \frac14 \int d^dx d^2\theta_1 e^{\e(\s_1 +\bar{\s}_1)} 
\left\{ G_{ij} D_{1\a} \Psi^i D_1^{\a} \Psi^j + \right. \nonumber \\
&~~& \left. + \e \left[ D_i K D_{1\a} \Psi^i D_1^{\a} 
(\s_1 + \bar{\s}_1)
+ iD_i K J^i_{~j} D_{1\a} \Psi^j D_1^{\a} (\s_1 - \bar{\s}_1) 
\right] + \right. \nonumber \\
&~~& \left. + \frac{\e^2}{2} ~K \left[ D_{1\a} (\s_1+\bar{\s}_1)D_1^{\a} (\s_1 +
\bar{\s}_1) - D_{1\a}(\s_1 -\bar{\s}_1) D_1^{\a}(\s_1-\bar{\s}_1) 
\right] \right\}
\label{5}
\ena
where $G_{ij}$ is the K\"ahler metric. 
We note that in a $(n+1)$--dimensional complex
manifold the model takes the familiar form
\EQ
S_{M} = \frac14 \int d^dx d^2 \theta_1 
{\cal G}_{IJ} D_{1\a}\x^I 
D_1^{\a} \x^J  
\EN
where we have defined $\x \equiv 
(\Psi^1,\cdots ,\Psi^{2n},\s_1 + \bar{\s}_1, \s_1 -\bar{\s}_1)$
and
\EQ
{\cal G}\equiv
\left( \begin{array} {ccc}
G_{ij} &  \frac{\e}{2} D_i K & i \frac{\e}{2} D_nK ~J^n_{~i} \\
\frac{\e}{2} D_j K & \frac{\e^2}{2} K & 0 \\
i \frac{\e}{2} D_mK ~J^m_{~j} & 0 & -\frac{\e^2}{2} K 
\end{array}
\right)e^{\e(\s_1 +\bar{\s}_1)}
\EN
It is easy to identify the terms which
correspond to dilaton--type
couplings. Indeed, integration by parts in (\ref{5}) leads to the
$N=1$ dilaton action
\bea
&& \int d^dx d^2 \theta_1 ~
\frac{\e}{2} \left[ \frac12 D_iK D_{1\a} \Psi^i D_1^{\a} (\s_1+\bar{\s}_1) 
+ \frac{\e}{4} K D_{1\a} (\s_1+\bar{\s}_1)D_1^{\a}(\s_1+\bar{\s}_1) \right]
e^{\e(\s_1+\bar{\s}_1)} = \nonumber \\
&&~~~~~~~~~~~~~~~~~~= \int d^dx d^2 \theta_1 
\frac{\e}{2} ~E_1^{-1}~R_1~ K
\label{6}
\ena
where $E_1 = e^{\frac{d-1}{2} (\s_1 +\bar{\s}_1)}$ is the $N=1$
vielbein superdeterminant and
the $N=1$ scalar curvature in $d= 2-2\e$ dimensions is given by \cite{b7}
\EQ
R_1 = \left[ (D_1)^2 (\s_1+\bar{\s}_1) - 
\frac{\e}{4} D_{1\a} (\s_1+\bar{\s}_1)
D_1^{\a} (\s_1+\bar{\s}_1)\right] e^{\frac12 (\s_1+\bar{\s}_1)} 
\EN
The remaining terms which contain 
$\s_1-\bar{\s}_1$, can be identified as the $N=2$ supersymmetric partners
of the dilaton couplings. In section 4 we will show that these vertices
give rise to divergent corrections to the metric $G_{ij}$, not expressible as 
geometric objects on the manifold. 

\sect{Quantization in superspace}

We study now the quantization of the system 
described by the action in (\ref{action})
directly in $N=2$ superspace.
For the bosonic case a covariant procedure of quantization for
the Hilbert--Einstein action away from two dimensions has been 
proposed in Ref. \cite {b10}. Within this approach it has been shown
that the coupling of nonconformal ($c\neq 0$) matter
to gravity leads to a
one--loop renormalization of the gravitational coupling constant 
$\k^2_0$.   As a consistency check of the procedure, the exact results 
of two--dimensional quantum gravity \cite{distler} have been reobtained for 
$\e \rightarrow 0$,
in the strong coupling regime ($\k^2 \gg |\e|$).
This analysis has been extended to the $N=1$ supersymmetric system \cite{b11}.

In the $N=2$ case one can use a parallel formulation. 
The renormalization of the gravitational coupling constant
is given by
\EQ
\frac{1}{\k^2_0}=\m^{-2\e} \left(\frac{1}{\k^2}-\frac{1}{2}
\frac{c-1}{\e}\right)
\label{ren}
\EN
Therefore, in complete analogy with the bosonic example
(see Ref. \cite{b5}, eq. (2.13)), 
we take as asymptotic behavior of $\k^2_0$
in the limits $c \rightarrow -\infty$, $\e \rightarrow 0$, and 
in the strong coupling regime 
\EQ
\k^2_0 \sim- \frac{2\e}{c}
\label{leadorder}
\EN
Alternatively this can be viewed as a definition of induced $d$-dimensional
gravity, to leading order in the anomaly coefficient $c$.

Our aim is to compute radiative gravitational corrections to the
$\b$--functions of the $N=2$ $\s$--model to leading order 
in the semiclassical limit $c\rightarrow -\infty$. Therefore
we consider contributions with only one $\k^2_0$
factor (see eq. (\ref{leadorder})). Since the dependence on the $H_a$
supergravity vector field has been dropped, we only need compute the $\s$
chiral compensator propagator and its couplings to the
$\Phi$ matter fields. Following a standard procedure, 
we use the quantum--background field method and
perform a linear splitting 
\EQ
\Phi^{\m} \rightarrow \Phi^{\m} + \Phi_0^{\m}
\qquad ~~~ \bar{\Phi}^{\bar{\m}} \rightarrow 
\bar{\Phi}^{\bar{\m}} +\bar{\Phi}_0^{\bar{\m}}
\EN
in the action (\ref{action}). With a rescaling $\e\s \rightarrow 
\s$, we obtain
\bea
S &=& \int d^dx d^2\theta d^2\bar{\theta} \left[ 1 +  
(\s+\bar{\s}) + \frac12  (\s+ \bar{\s})^2 + \cdots \right]
\nonumber \\
&~& \left[-2 \frac{d-1}{d-2} \k_0^{-2}
+K(\Phi_0,\bar{\Phi}_0) + K_{\m}(\Phi_0,\bar{\Phi}_0)
\Phi^{\m} + K_{\bar{\m}}(\Phi_0,\bar{\Phi}_0) 
\bar{\Phi}^{\bar{\m}} + \right. \nonumber \\
&~~~& \left. 
+ \frac12 K_{\m \n}(\Phi_0,\bar{\Phi}_0) \Phi^{\m} \Phi^{\n}
+ \frac12 K_{\bar{\m} \bar{\n}}(\Phi_0,\bar{\Phi}_0)
\bar{\Phi}^{\bar{\m}} \bar{\Phi}^{\bar{\n}} +
K_{\m \bar{\n}}(\Phi_0,\bar{\Phi}_0) \Phi^{\m} 
\bar{\Phi}^{\bar{\n}}+ \cdots \right]
\label{7}
\ena
where $\Phi^{\m}$, $\bar{\Phi}^{\bar{\m}}$, $\s$, $\bar{\s}$
are the quantum fields. 
From (\ref{7}) we have the supergravity propagator 
in the usual form for chiral superfields
(in our conventions $\Box = \frac12 \pa^a \pa_a$)
\EQ
\langle \s(x,\theta,\bar{\theta}) 
\bar{\s}(x',\theta',\bar{\theta}')  \rangle
= -\frac{\k_0^2 }{d-1} \e ~\Box^{-1} \d^{(2)} (x-x') ~\d^{(2)} (\theta-\theta') 
~\d^{(2)} (\bar{\theta} -\bar{\theta}')
\label{7bis}
\EN
Since we have performed a linear
splitting the various terms in the expansion (\ref{7}) are not 
manifestly covariant under reparametrization of the manifold. 
However, in the final result the local 
divergent contributions to the effective action, and correspondingly 
the counterterms, will be in covariant form \cite{b8}.

In Ref. \cite{b8} where the flat (i.e. in the absence of
supergravity) $N=2$ $\s$--model was
studied, a general procedure for the perturbative evaluation
of the ultraviolet divergences was outlined, which  
on the basis of dimensional considerations and $N=2$ supersymmetry,
lead to simplified Feynman rules. 
In particular the relevant contributions to the tree--level
two--point function $\langle\Phi^{\m} \bar{\Phi}^{\bar{\m}}
\rangle$  were computed to all orders in the background,
and an effective matter propagator was obtained 
\EQ
\langle \Phi^{\m}(x,\theta,\bar{\theta}) 
\bar{\Phi}^{\bar{\n}}(x',\theta',\bar{\theta}')  \rangle
= -K^{\m \bar{\n}} \Box^{-1} \d^{(2)} (x-x') ~\d^{(2)} (\theta-\theta') 
~\d^{(2)} (\bar{\theta} -\bar{\theta}')
\label{8}
\EN
where $K^{\m \bar{\n}}$ is the inverse of the Kahler metric. 
In addition a simplified set of effective vertices
was introduced: indeed it was shown
that Feynman diagrams containing  
vertices of the form $K_{\m \n \rho \dots}$ and/or 
$K_{\bar{\m} \bar{\n} \bar{\rho} \dots}$ with only unbarred or
only barred indices do not contribute to divergent quantum
corrections. 
Therefore, in the case of flat $N=2$ superspace 
the counterterms are always in terms of derivatives of the K\"ahler
metric and expressible as geometric objects (products of the Riemann
tensor and its derivatives) in the final result.

In the presence of propagating gravity fields, the same 
dimensional arguments do apply and 
exactly the same conclusions can be drawn as far as
the matter effective propagator and the matter effective 
self--interactions are concerned. 
On the other hand, whenever matter--supergravity couplings 
are involved some care is needed in order to identify 
correctly the relevant ones. In this case one has to keep vertices
where at least one quantum field has opposite chirality with
respect to the others. Therefore, we cannot 
discard vertices with only unbarred or barred indices on $K$ if
a quantum gravity line of opposite chirality is present, i.e.
vertices like $K_{ \m \n \dots\rho } \Phi^\m \Phi^\n\dots \Phi^{\rho} \bar{\s}$
are {\em relevant}.

With this set of rules in mind we proceed as follows:
first, since we study supergravity
effects in the semiclassical limit $c \rightarrow -\infty$,
at any loop order we draw all the diagrams which have only one 
gravity line (the leading order in $1/c$). 
Then on each diagram we perform the $D$--algebra as explained   
in Ref. \cite{b8} and reduce the corresponding expressions
to standard momentum integrals which we evaluate using supersymmetric
dimensional regularization and minimal subtraction. In order to extract the 
overall divergence we subtract the ultraviolet
subdivergences corresponding to lower--order renormalizations 
and remove infrared infinities by using 
the procedure described in Refs. \cite{b12,b13,b5}. 

The quantum counterterms we want to compute correspond to local 
corrections to the 
K\"ahler potential. In dimensional regularization they have the form
\EQ
K \rightarrow K + \sum_{n=1}^{\infty} \sum_{l=n}^{\infty}
\frac{1}{\e^n} K^{(n,l)}
\label{9}
\EN
so that the renormalization of the  
K\"ahler metric is given by
\EQ
G_{\m \bar{\n}}^B = G_{\m \bar{\n}}^R+ \sum_{n=1}^{\infty} 
\sum_{l=n}^{\infty} \frac{1}{\e^n} T_{\m \bar{\n}}^{(n,l)} 
\EN
Correspondingly the $\b$--function is 
\EQ
\b_{\m \bar{\n}}(G^R) = 2\e G_{\m \bar{\n}}^R + 2\left( 1+ 
\l \frac{\pa}{\pa \l} \right) \sum_{l=1}^{\infty} 
T_{\m \bar{\n}}^{(1,l)} (\l^{-1} G^R)|_{\l=1}
\label{9bis}
\EN
Therefore in order to evaluate the perturbative  
corrections to the $\b$--functions 
we concentrate 
on  the first order pole contributions in the $\e$--expansion.
This allows to discard Feynman diagrams  
 which reduce to tadpole--like
diagrams when in the process of 
performing the $D$-algebra, matter propagators are cancelled
by momentum factors. Indeed it is easy to show that
after sutraction of subdivergences they give rise
to higher order $1/\e$ poles. On the contrary,
a careful analysis of the subraction of subdivergences
shows that in general we cannot drop 
corresponding diagrams in which
a gravity propagator would be cancelled. We illustrate
this rather subtle point in Appendix B with a specific example.

As it has been observed in Ref. \cite{b3}, for the $N=2$ theories 
the cosmological term is given by a chiral integral and as such
it is not renormalized. Consequently the  Liouville field does not
acquire anomalous dimensions and the physical scale is not modified
with respect to the standard renormalization scale.
This implies that the renormalization group $\b$--functions 
in the presence of $N=2$ dynamical gravity are not rescaled by
a multiplicative factor.
The only possible corrections, if present at all, must be given by 
new, nontrivial contributions from gravity propagating inside the loops.
At one loop no modification has been found \cite{b3}.
  
In the next section we present the explicit calculation of
the gravitational dressing of the $\b$--function 
at two loops in the matter fields.

\sect{Gravitational dressing at two--loop order}

Now we consider quantum corrections at two loops in the presence of gravity.
(Note that whenever we say at $n$--loop order in the matter
fields we are considering $n$--loop diagrams with a gravity
propagator inserted, so that we are effectively at $n+1$ loops.) 
Since the $\s$ propagator is $O(\e^2)$, the first 
divergent, gravitational--induced contribution 
 can only arise at two loops in the matter fields. 

In the absence of supergravity the second 
order correction to the K\"ahler potential has been computed 
in Ref. \cite{b8} and it is given by $K^{(2,2)} = R$. Therefore, 
as it is well known, in the flat case the $\b$--function does not receive
any contribution at two loops.

In order to evaluate the gravitational dressing
we need consider two-loop matter graphs with one
$\s$ propagator inserted. Keeping in mind the set of effective
Feynman rules discussed in the previous section, one selects the
relevant diagrams which in the end will give contributions
proportional to $1/\e$. They are drawn in Figs.
1, 2, 3 where the structure of the $D$--derivatives 
is explicitly indicated. We have not drawn diagrams containing
two--point matter--gravity vertices that by integration by parts of 
$D^2$ factors reduce immediately to the structures shown
in the figures. Their dependence on the background contributes to
covariantize the 
matter couplings 
according to the formulae (\ref{a20}--\ref{a22}) (see Appendix A 
for more details). 

We perform the $D$ algebra in such a way to reduce all the
diagrams in Figs. 1, 2, 3 to the diagrams $1a$, $2a$ and $3a$ 
respectively. Again this can be achieved by partial integration:
at the vertices which contain only one $D^2$ ($\bar{D}^2$)
and a number of $\bar{D}^2$ ($D^2$) we integrate by parts
one $\bar{D}^2$ ($D^2$) and use the relations 
$ \bar{D}^2 D^2 \bar{D}^2 = \Box \bar{D}^2 $, 
$D^2 \bar{D}^2 D^2  = \Box D^2$ to cancel the propagator
of the corresponding line. Then the $\bar{D}^2$
($D^2$) factor is integrated back to the original line. 
Applying this procedure a number of times one obtains the above 
stated result.  The dependence on the background fields can be 
reconstructed diagram by diagram looking at the structure of the
vertices, so that including combinatoric factors and making use of
the relations (\ref{a19}, \ref{a20}--\ref{a22}) we obtain the 
following result: from the diagrams in Fig.1 
\EQ
\frac{1}{6} D_{\m} D_{\n} D_{\rho} K~ D^{\m} D^{\n} D^{\rho} K 
~I_1
\label{10}
\EN
where $I_1$ is the contribution corresponding to diagram $1a$.
Diagrams in Fig. 2 sum up to 
\EQ
\frac{1}{2} D_{\bar{\m}} D_{\n} D_{\rho} K~ 
D^{\bar{\m}} D^{\n} D^{\rho} K ~I_2
\label{11}
\EN
where $I_2$ corresponds to the diagram $2a$, whereas
from the diagrams in Fig.3 we obtain 
\EQ
\frac{1}{4} R_{\m \bar{\m} \n \bar{\n}} ~D^{\m} D^{\n} K 
~D^{\bar{\m}} D^{\bar{\n}} K ~I_3
\label{12}
\EN
$I_3$ being the contribution of diagram $3a$.
To evaluate $I_1$, $I_2$ and $I_3$ we first complete the
$D$--algebra as indicated in Figs. 4, 5, 6.
Whenever we produce a $\Box$ factor on a matter line we drop the
corresponding tadpole contribution.  
Moreover we integrate $\pa$--derivatives by parts
in order to reduce all the contributions to products of
tadpoles that, as explained in the previous section and in 
Appendix B, we keep only if the cancelled propagator is a gravity
line. 
In dimensional regularization and in our IR subtraction 
scheme \cite{b12,b5}, the elementary tadpole integral 
$I= \int \frac{d^dp}{(2\p)^d}\frac{1}{p^2}$ is computed by shifting 
the propagator $\frac{1}{p^2} \rightarrow \frac{1}{p^2} +
\frac{\p}{\e} \d^{(2)}(p^2) $, so that one has
\EQ
I \equiv \int \frac{d^dp}{(2\p)^d} \frac{1}{p^2} \rightarrow 
\frac{1}{4\pi} \frac{1}{\e} 
\label{13}
\EN
In this fashion the overall divergence of the diagrams in Figs. 4, 5, 6  
can be easily determined 
\bea
I_1 &=& -\k_0^2 \e I^3 = \frac{1}{(4\p)^3}\frac{2}{c}
\frac{1}{\e} \nonumber \\
I_2 &=& \k_0^2\e(- I^3 +\frac{2}{3} I^3) = \frac{1}{(4\p)^3}\frac{2}{3c}
\frac{1}{\e}  \nonumber \\
I_3 &=&\k_0^2\e \frac{4}{3} I^3 = - \frac{1}{(4\p)^3}\frac{8}{3c}
\frac{1}{\e}
\label{14}
\ena
Collecting the results from eqs. (\ref{10}--\ref{12})
and (\ref{14}) we obtain the final expression for
the $1/c$--correction to the K\"ahler potential at
two--loop order 
\bea
K^{(1,2)} &=& \frac{1}{(4\p)^3}\frac{1}{3c}  \left[ D_{\m} D_{\n} 
D_{\rho} K~ D^{\m} D^{\n} D^{\rho} K  + D_{\bar{\m}} D_{\n} 
D_{\rho} K~ D^{\bar{\m}} D^{\n} D^{\rho} K  \right. \nonumber \\
&~&~~~~~~~~~~\left. - 2 R_{\m \bar{\m} \n \bar{\n}} D^{\m} D^{\n} K
D^{\bar{\m}} D^{\bar{\n}} K \right]
\label{17}
\ena
The result can be expressed in terms of real coordinates as
\EQ
K^{(1,2)} = \frac{1}{(4\p)^3}\frac{1}{3c} \left[ \frac12 D_i D_j D_l
K~ D^i D^j D^l K - R_{ijlm} D^i D^l K~ D^j D^m K - 2 R 
\right] 
\label{18}
\EN
From the expression (\ref{17}) we immediately obtain the
$1/\e$--correction to the K\"ahler metric 
\EQ
T_{\m \bar{\n}}^{(1,2)} = \frac{\pa K^{(1,2)}}{\pa \Phi^{\m} 
\pa \bar{\Phi}^{\bar{\n}}}
\label{19}
\EN
We conclude that in the presence of propagating gravity 
the matter $\b$--function is corrected at two loops  
\bea
\b_{\m \bar{\n}}^{(2)} &=& \frac{1}{(4\p)^3}\frac{2}{3c}~D_{\m} D_{\bar{\n}}
\left[ D_{\rho} D_{\s} 
D_{\t} K~ D^{\rho} D^{\s} D^{\t} K  + D_{\bar{\rho}} D_{\s} 
D_{\t} K~ D^{\bar{\rho}} D^{\s} D^{\t} K  \right. \nonumber \\
&~~&~ \left. - 2 R_{\rho \bar{\rho} \s \bar{\s}} D^{\rho} D^{\s} K
D^{\bar{\rho}} D^{\bar{\s}} K \right]
\label{20}
\ena
In order to reexpress the above result in terms of real coordinates 
we need introduce the complex structure $J^i_{~j}$. We have
\EQ
\b^{(2)}_{ij} =  (D_i D_j + J^m_{~i} J^n_{~j}  
D_m D_n) K^{(1,2)}
\label{21}
\EN
where $K^{(1,2)}$ is given in Eq. (\ref{18}).

The presence of the complex structure in the metric 
$\b$--function is consistent with the fact that 
in the $N=1$ formalism the gravity--matter coupling  
in the bare action (\ref{5}) is indeed proportional to $J^i_{~j}$. 

\sect{Conclusions}
The main result of our work is contained in eq. (\ref{20}): the
 $\b$--function of the $N=2$ supersymmetric $\s$--model
receives its first gravitational correction at two loops in the  
matter fields. 
The dressing, absent at lower orders, is given by structures which are 
{\em not}
geometric objects of the K\"ahler manifold.
This new type of divergences can be interpreted as dilatonic
contributions. As mentioned earlier, once the interaction between 
the K\"ahler potential and $N=2$ supergravity is switched on,
supergravity--dilaton vertices are automatically
included: they vanish in the classical limit, being
 $O(\e)$, but at the quantum level a dilaton term is induced in
the effective action. 
At two loops in the matter system with one insertion of $N=2$ 
supergravity, it is the dilaton which gives a nonvanishing
correction to the metric $\b$--function.

We observe that our calculation is not affected by scheme 
dependence ambiguities.
In fact the first, nonvanishing, 
gravitational correction is the one in eq. (\ref{20}). 
Therefore no $O(1/c)$ conventional subtraction 
ambiguities can be produced from finite
subtractions proportional to lower--loop counterterms. Moreover,
even if one were to be perverse and take into account finite subtractions
proportional to the one--loop matter counterterm, the ambiguities would be
given by geometric structures and they would never mix with the 
terms in eq. (\ref{20}).
 
In order to fully understand the role played by the dilaton field,
it would be quite interesting to consider the $N=1$ supersymmetric 
$\s$--model
with both metric and dilaton couplings to dynamical supergravity. 
For this system one could
compute  the corrections to the metric $\b$--function
induced by the supergravity--dilaton interactions and interpret then 
the equations $\b_{ij}=0$ as equations of motion for the
noncritical superstring.

\vskip 20pt
{\bf Acknowledgements}: This work has been partially supported by
INFN and the European Commission TMR program ERBFMRX--CT96--0045 in 
which Milano is associated to Torino.

\vskip 20pt
\appendix
\sect{K\"ahler manifolds}

A K\"ahler manifold is a complex manifold with vanishing torsion. 
It is endowed with a complex structure $J^i_{~j}$ which satisfies
$J^2=-1$ and is an isometry of the metric
\EQ
J^i_{~j} J^k_{~l} ~G_{ik} = G_{jl}
\label{a1}
\EN
Moreover it is covariantly constant as a consequence of 
the vanishing of the torsion.  
Therefore it is always possible to choose 
a suitable set of complex coordinates on the manifold 
so that the complex structure has the standard form
\EQ
J^{\m}_{~\n} = i \d^{\m}_{~\n} \qquad \qquad 
J^{\bar{\m}}_{~\bar{\n}} = -i \d^{\bar{\m}}_{~\bar{\n}}
\label{a2} 
\EN
In this coordinate system the K\"ahler metric satisfies 
\EQ
\pa_{\rho} G_{\m \bar{\n}} = \pa_{\m} G_{\rho \bar{\n}}
\qquad  
\pa_{\bar{\rho}} G_{\m \bar{\n}} = \pa_{\bar{\n}} G_{\m \bar{\rho}}
\qquad G_{\m \n} = 0
\EN
and locally it can be expressed in terms of the K\"ahler 
potential $K$ as
\EQ
G_{\m \bar{\n}} = \frac{\pa K }{\pa \Phi^{\m} \pa \bar{\Phi}^{\bar{\n}}}
\EN
In general we use the following notation
\EQ
K_{\m_1 \cdots \m_p \bar{\n}_1 \cdots \bar{\n}_q} \equiv
\frac{\pa^p~~~~~~~~~~}{\pa \Phi^{\m_1} \cdots \pa \Phi^{\m_p}}
\frac{\pa^q~~~~~~~~~~}{\pa \bar{\Phi}^{\bar{\n}_1} \cdots \pa 
\bar{\Phi}^{\bar{\n}_q}} K
\EN
so that $G_{\m \bar{\n}} = K_{\m \bar{\n}}$.

The only nonvanishing components of the connection are
$\G^{\m}_{\n \rho}$ and $\G^{\bar{\m}}_{\bar{\n} \bar{\rho}}$, 
and the Riemann and Ricci tensors are given by 
\EQ
R_{\m \bar{\m} \n \bar{\n}} =  K_{\m \n \bar{\m} \bar{\n}}
- K_{\m \n \bar{\rho}} K_{\rho \bar{\m} \bar{\n}} K^{\rho \bar{\rho}}
\label{a19}
\EN
\EQ
R_{\m \bar{\n}} \equiv R^{\rho}_{~\m \bar{\n} \rho} = \pa_{\m}
\pa_{\bar{\n}} \log {\rm {det}} K_{\s \bar{\s}}
\EN
We list some useful identities involving covariant derivatives 
of the K\"ahler potential which have been used in the
calculation of the corrections to the metric $\b$--function
\EQ
D_{\m} D_{\n} K = K_{\m \n} - K_{\m \n \bar{\rho}} K^{\rho \bar{\rho}}
K_{\rho}
\label{a20}
\EN
\EQ
D_{\bar{\m}} D_{\n} D_{\rho} K = \left[ -K_{\bar{\m} \n \rho \bar{\s}}
+ K_{\bar{\eta} \n \rho} K_{\bar{\m} \eta \bar{\s}} K^{\eta
\bar{\eta}} \right] K^{\s \bar{\s}} K_{\s}                        
\label{a21}
\EN
\EQ
D_{\m} D_{\n} D_{\rho} K = K_{\m \n \rho} - K^{\s \bar{\s}} \left[
K_{\m \n \rho \bar{\s}} K_{\s} -3 K_{\bar{\s} (\m \n} 
K_{\rho )\s} + 3 K^{\eta \bar{\eta}} K_{\bar{\s} (\m \n} 
K_{\rho )\s \bar{\eta}} K_{\eta} \right]
\label{a22}
\EN
At the perturbative level these covariant couplings come from 
resummation of different contributions involving mixed matter--gravity
vertices. As an example we consider a vertex proportional to $K_{\m \n}$
with two matter and one 
gravity lines  as in Fig. $7a$. 
As shown in Fig. $7b$ a graph
with a mixed two--point vertex reduces to the one in Fig. $7a$ once 
the $D^2$ on the gravity line has been integrated by parts on the matter.
Its background structure $-K_{\m \n \bar{\rho}} K^{\rho \bar{\rho}}
K_{\rho}$ is indeed what one needs in order to covariantize 
$K_{\m \n}$ as in Eq. (\ref{a20}).

\sect{An example}

In this appendix we show on a simple example how one must carefully
operate when momentum factors, which cancel the gravity propagator,
are produced by $D$--algebra manipulations.

To be pedagogical let us consider first the graph in Fig. $8a$ where the
$D^2$ factors are explicitly indicated at one vertex.
By performing the $D$--algebra the matter propagator 
carrying the $\bar{D}^2$ factor can be immediately
cancelled and we are left with the momentum structure shown in Fig. $8b$.
This graph is tadpole--like and does not contribute to the $1/\e$
pole. In fact, in the evaluation of the corresponding integral we must
subtract two one--loop subdivergences (corresponding to the $A$ and $B$
loops),
and one two--loop subdivergence (corresponding to the subgraph $A\cup B$) 
and in so doing we obtain 
\EQ
(\e^2 I)I^2-2\frac{1}{\e}(\e^2I)I
-\left[ I^2-2\frac{1}{\e}I\right]_{div}(\e^2I)=0
\EN
where $I$ is the elementary tadpole integral (\ref{13}) and we have 
neglected
$1/4\pi$--factors for notational convenience.

Let us consider now the graph in Fig. $8c$.
In this case by performing the $D$--algebra the gravity propagator 
is cancelled and the resulting momentum structure is shown in Fig. $8d$.
Again we are lead to evaluate a tadpole--like integral, 
but now none of the subgraphs
is divergent because either it does not include the (cancelled) gravity
line and then it is finite by itself, or else it includes the cancelled
gravity and then it becomes finite once multiplied by
 the $\e^2$ carried by the gravity propagator.
Hence there are no subdivergences and the result
is simply
\EQ
(\e^2 I) I^2=\frac{1}{\e}
\EN

More generally, when the gravity propagator has been cancelled
we can obtain contributions of order $1/\e$ even if the
graph corresponding to the momentum structure is tadpole--like.
In these cases we cannot discard the graph.

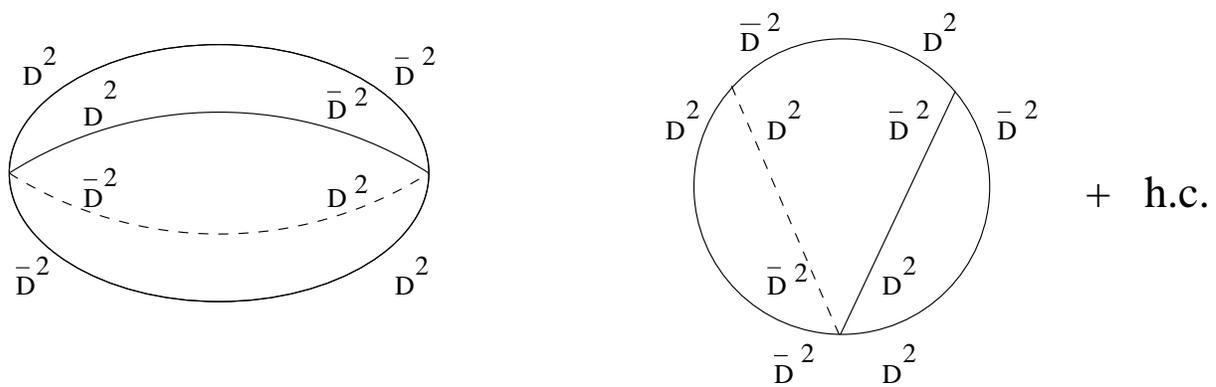
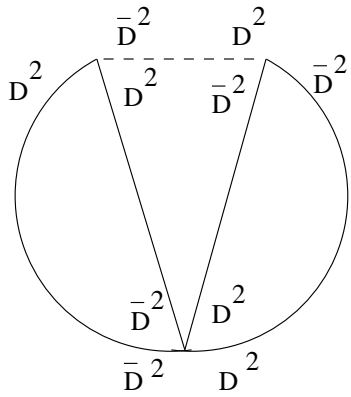
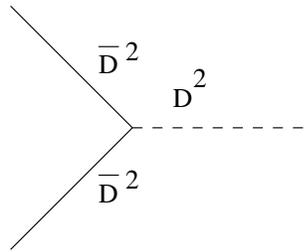
\end{document}